\documentclass[nonacm]{acmart}
\usepackage[utf8]{inputenc}

\usepackage{babel}
\usepackage{multirow}
\usepackage[inline]{enumitem}
\usepackage{caption}
\usepackage{moresize}
\usepackage{nicefrac}

\usepackage{multicol}
\usepackage[ruled, linesnumbered, longend, lined, nofillcomment]{algorithm2e} 
\newcommand{\algoFontSize}{small}

\SetCommentSty{mycommentfont}

\usepackage{tikz}
\usetikzlibrary{shapes,arrows,positioning,decorations.markings}

\newcommand{\JJJ}{Jayanti, Jayanti and Joshi}
\newcommand{\bigO}[1]{\mathcal{O}(#1)}
\newcommand{\bigOmega}[1]{{\Omega}(#1)}

\newcommand{\broadcast}{\emph{Broadcast}}
\newcommand{\bSet}{\textsc{Set}}
\newcommand{\bWait}{\textsc{Wait}}
\newcommand{\bRead}{\textsc{Read}}

\newcommand{\newnode}{new\_node}
\newcommand{\retire}{retire\_last\_node}

\newcommand{\NCS}{\texttt{NCS}}
\newcommand{\CS}{\texttt{CS}}
\newcommand{\Recover}{\texttt{Recover}}
\newcommand{\Enter}{\texttt{Enter}}
\newcommand{\Exit}{\texttt{Exit}}
\newcommand{\segment}{segment}

\newcommand{\Freed}{\textsc{Free}}
\newcommand{\Allocated}{\textsc{Allocated}}
\newcommand{\Retired}{\textsc{Retired}}
\newcommand{\Reclaimed}{\textsc{Reclaimed}}

\title{Memory Reclamation for Recoverable Mutual Exclusion}
\author{Sahil Dhoked}
\affiliation{
  \institution{The University of Texas at Dallas}
  \state{TX}
  \postcode{75080}
  \country{USA}}
\email{sahil.dhoked@utdallas.edu}
\author{Neeraj Mittal}
\affiliation{
  \institution{The University of Texas at Dallas}
  \state{TX}
  \postcode{75080}
  \country{USA}}
\email{neerajm@utdallas.edu}
\acmBooktitle{}

\begin{document}
    \begin{abstract}
        Mutual exclusion (ME) is a commonly used technique to handle conflicts in concurrent systems. With recent advancements in non-volatile memory technology, there is an increased focus on the problem of recoverable mutual exclusion (RME), a special case of ME where processes can fail and recover. However, in order to ensure that the problem of RME is also of practical interest, and not just a theoretical one, memory reclamation poses as a major obstacle in several RME algorithms. Often RME algorithms need to allocate memory dynamically, which increases the memory footprint of the algorithm over time. These algorithms are typically not equipped with suitable garbage collection due to concurrency and failures.
        
        In this work, we present the first ``general'' recoverable algorithm for memory reclamation in the context of recoverable mutual exclusion. Our algorithm can be plugged into any RME algorithm very easily and preserves all correctness property and most desirable properties of the algorithm. The space overhead of our algorithm is $\mathcal{O}(n^2 * sizeof(node)\ )$, where $n$ is the total number of processes in the system. In terms of remote memory references (RMRs), our algorithm is RMR-optimal, \emph{i.e}, it has a constant RMR overhead per passage. Our RMR and space complexities are applicable to both \textit{CC} and \textit{DSM} memory models.
    \end{abstract}
    \maketitle
    
    \section{Introduction}
    
    Mutual exclusion (ME) is a commonly used technique to handle conflicts in concurrent systems. The problem of mutual exclusion was first defined by Dijkstra \cite{Dij:1965:CACM} more than half a century ago. Mutual exclusion algorithms, commonly known as locks, are used by processes to execute a part of code, called critical section (CS) in isolation without any interference from other processes. The CS typically consists of code that involves access to shared resources, which when accessed concurrently could potentially cause undesirable race conditions. The mutual exclusion problem involves designing algorithms to ensure processes enter the CS one at a time.
    
    Generally, algorithms for mutual exclusion are designed with the assumption that failures do not occur, especially while a process is accessing a lock or a shared resource. However, such failures can occur in the real world. A power outage or network failure might create an unrecoverable situation causing processes to be unable to continue. If such failures occur, traditional mutual exclusion algorithms, which are not designed to operate properly under failures, may deadlock or otherwise fail to guarantee important safety and liveness properties. In many cases, such failures may have disastrous consequences. This gave rise to the problem of \emph{recoverable mutual exclusion (RME)}. The RME problem involves designing an algorithm that ensures mutual exclusion under the assumption that process failures may occur at \emph{any} point during their execution, but the system is able to recover from such failures and proceed without any adverse consequences.
    
    Traditionally, concurrent algorithms use checkpointing and logging to tolerate failures by regularly saving relevant portion of application state to a persistent storage such as hard disk drive (HDD). Accessing a disk is orders of magnitude slower than accessing main memory. As a result, checkpointing and logging algorithms are often designed to minimize disk accesses.
    \emph{Non-volatile random-access memory (NVRAM)} is a new class of memory technologies that combines the low latency and high bandwidth of traditional random access memory with the density, non-volatility, and economic characteristic of traditional storage media (\emph{e.g.}, HDD). Existing checkpointing and logging algorithms can be modified to use NVRAMs instead of disks to yield better performance, but, in doing so,we would not be leveraging the true power of NVRAMs \cite{NarHod:2010:ASPLOS, GolRam:2019:DC}. NVRAMs can be used to directly store implementation specific variables and, as such, have the potential for providing near-instantaneous recovery from failures. 
    
    By directly storing implementation variables on NVRAMs, most of the application data can be easily recovered after failures. However, recovery of implementation variables alone is not enough. Processor state information such as contents of program counter, CPU registers and execution stack cannot be recovered completely and need to be handled separately. Due to this reason, there is a renewed interest in developing fast and dependable algorithms for solving many important computing problems in software systems vulnerable to process failures using NVRAMs. Using innovative methods, with NVRAMs in mind, we aim to design efficient and robust fault-tolerant algorithms for solving mutual exclusion and other important concurrent problems.
    
    The RME problem in the current form was formally defined a few years ago by Golab and Ramaraju in \cite{GolRam:2016:PODC}. Several algorithms have been proposed to solve this problem \cite{GolRam:2019:DC,GolHen:2017:PODC,JayJos:2017:DISC,JayJay+:2019:PODC,DhoMit:2020:PODC}. However, in order to ensure that the problem of RME is also of practical interest, and not just a theoretical one, memory reclamation poses as a major obstacle in several RME algorithms. Often, RME algorithms allocate memory dynamically which increases the memory footprint of the algorithm over time. These algorithms are typically not equipped with suitable garbage collection to avoid errors that may arise from concurrency and potential failures.
    
    Memory reclamation, in single process systems without failures, follows a straightforward pattern. The process allocates ``nodes'' dynamically, consumes it, and frees it once it has no more need of this node. Freed nodes may later be reused (as part of a different allocation) or returned to the operating system. However, due to some programmer error, if a node that is freed is later accessed by the process in the context of the previous allocation, it may cause some serious damage to the program and the operating system as well. In the context of multi-process systems, when a process frees a node, we may face the same issue without any programmer error. Even if the process that frees the node is able to guarantee that it will not access that node again, there may exist another process that is just about to access or dereference the node in the context of the old allocation.
    
    In order to avoid the aforementioned error, freeing a node is broken down into two tasks. First, a process retires the node, after which, any process that did not have access to the node may no longer be able to get access to the node. Second, the node needs to be reclaimed once it is deemed to be ``safe'', \emph{i.e.}, no process can obtain any further access to the node in the context of the previous allocation. A memory reclamation service is responsible to provide a safe reclamation of a node once it is retired. On the other hand, the responsibility of retiring the node is typically on the programmer that needs to consume the memory reclamation service.
    
    Prior works on memory reclamation \cite{Mic:2004:TPDS, Fra:2004:PhD, mckenney1998read, arcangeli2003using, Bro:2015:PODC, wen:2018:ppopp} provide safe memory reclamation in the absence of failures, but are not trivially suited to account for failures and subsequent recovery using persistent memory. Moreover, most works focus on providing memory reclamation in the context of lock-free data structures.
    
    In this work, we present the first ``general'' recoverable algorithm (that we know of) for memory reclamation in the context of recoverable mutual exclusion. Our algorithm is general enough that it can be plugged into any RME algorithm very easily, while preserving all correctness properties and most desirable properties of the algorithm. On the other hand, it is specific enough to take advantage of assumptions made by RME algorithms. In particular, our algorithm may be blocking, but it is suitable in the context of the RME due to the very blocking nature of the RME problem.
    
    Our approach derives from prior works of EBR \cite{Fra:2004:PhD} (epoch based reclamation) and QSBR \cite{mckenney1998read} (quiescent state based reclamation). However, unlike EBR and QSBR, where the memory consumption may grow unboundedly due to a slow process, our algorithm guarantees a bounded memory consumption. The space overhead of our algorithm is $\mathcal{O}(n^2 * sizeof(node)\ )$, where $n$ is the total number of processes in the system, and a ``node'' is a collection of all the resources used in one passage of the CS.
    
    One of the most important measures of performance of an RME algorithm is the maximum number of \emph{remote memory references (RMRs)} made by a process per critical section request in order to acquire and release the lock as well as recover the lock after a failure. Whether or not a memory reference incurs an RMR depends on the underlying memory model. The two most common memory models used to analyze the performance of an RME algorithm are \emph{cache-coherent (CC)} and \emph{distributed shared memory (DSM)} models. In terms of remote memory references (RMRs), our algorithm is RMR-optimal, \emph{i.e}, it has a constant RMR overhead per passage for both \textit{CC} and \textit{DSM} memory models. Moreover, this algorithm uses only read, write and comparison based primitives, 
    
    The main idea behind our approach is
    \begin{enumerate*}
        \item maintain two pools of ``nodes'', clean (reclaimed) and dirty (retired)
        \item wait for dirty nodes to become clean, while consuming the clean pool
        \item switch dirty and clean pools
    \end{enumerate*}.
    Our algorithm operates in tandem with any RME algorithm via two methods/APIs that can be invoked by the programmer to allocate new nodes and retire old nodes.
    
    \paragraph{Roadmap:}
    The rest of the text is organized as follows. We describe our system model and formally define the RME and the memory reclamation problem in \autoref{sec:model|problem}. 
    We define a new object, called the \broadcast{} object and its properties in \autoref{sec:broadcast}. We also present an RMR-optimal solution to the \broadcast{} object for both the CC and DSM model in \autoref{sec:broadcast}.
    In \autoref{sec:mem_rec}, we present an algorithm that provides memory reclamation for RME algorithms. This algorithm is RMR-optimal, but not lock-free. In \autoref{sec:application}, we describe how our memory reclamation algorithm can be equipped to existing RME algorithms. A detailed description of the related work is given in \autoref{sec:related}.
    Finally, in \autoref{sec:concl_future}, we present our conclusions and outline directions for future research.
    
    \section{System Model and Problem Formulation}
    \label{sec:model|problem}
        We assume that RME algorithms follow the same model and formulation as used by Golab and Ramaraju \cite{GolRam:2019:DC}. 
        
    \subsection{System model}
        We consider an asynchronous system of $n$ processes ($p_1, p_2, \dots, p_n$). Processes can only communicate by performing read, write and read-modify-write (RMW) instructions on shared variables. Besides shared memory, each process also has its private local memory that stores variables only accessible to that process (\emph{e.g.}, program counter, CPU registers, execution stack, \emph{etc.}).  Processes are not assumed to be reliable and may fail.
        
        A system execution is modeled as a sequence of process steps. In each step, some process either performs some local computation affecting only its private variables or executes one of the available instructions (read, write or RMW) on a shared variable or fails. Processes may run at arbitrary speeds and their steps may interleave arbitrarily. In any execution, between two successive steps of a process, other processes can perform an unbounded but finite number of steps.
        
        To access the critical section, processes synchronize using a recoverable \emph{lock} that provides mutual exclusion (ME) despite failures.
        
   \subsection{Failure model}
        We assume the \emph{crash-recover} failure model. A process may fail at any time during its execution by crashing. A crashed process recovers eventually and restarts its execution from the beginning. A crashed process does not perform any steps until it has restarted. A process may fail multiple times, and multiple processes may fail concurrently.
        
        
        On crashing, a process loses the contents of all volatile private variables, including but not limited to the contents of its program counter, CPU registers and execution stack. However, the contents of the shared variables and non-volatile private variables remain unaffected and are assumed to persist despite any number of failures. When a crashed process restarts, all its volatile private variables are reset to their initial values.
        
        Processes that have crashed are difficult to distinguish from processes that are running arbitrarily slow. However, we assume that every process is live in the sense that a process that has not crashed eventually executes its next step and a process that has crashed eventually recovers. In this work, we consider a failure to be associated with a single process.

    \subsection{Process execution model}
        The process execution for RME algorithms is modeled using two types of computations, namely \emph{non-critical section} and \emph{critical section}. A critical section refers to the part of the application program in which a process needs to access shared resources in isolation. A non-critical section refers to the remainder of the application program. 
       
        
		\begin{algorithm}[t]
			\begin{\algoFontSize}
				\DontPrintSemicolon
				\While {true} 
				{
					Non-Critical Section (NCS)\;
					Recover\;
					Enter\;
					Critical Section (CS)\;
					Exit\;
				}
				\caption{Process execution model}
				\label{algo:PEM}
			\end{\algoFontSize}
		\end{algorithm}
        
        The execution model of a process with respect to a lock is depicted in Algorithm~\ref{algo:PEM}. As shown, a process repeatedly executes the following five  \segment{s} in order: \NCS{}, \Recover{}, \Enter{}, \CS{} and \Exit{}.
        The first \segment{}, referred to as \NCS, models the steps executed by a process in which it only accesses variables outside the lock.
        The second \segment{}, referred to as \Recover, models the steps executed by a process to perform any cleanup required due to past failures and restore the internal structure of the lock to a consistent state. 
        The third \segment{}, referred to as \Enter, models the steps executed by a process to acquire the lock so that it can execute its critical section in isolation.
        The fourth \segment{}, referred to as \CS, models the steps executed by a process in the critical section where it accesses shared resources in isolation.
        Finally, the fifth \segment{}, referred to as \Exit, models the steps executed by a process to release the lock it acquired earlier in \Enter{} \segment. 
       
        It is assumed that in the \NCS{} \segment{}, a process does not access any part of the lock or execute any computation that could potentially cause a race condition. Moreover, in \Recover{}, \Enter{} and \Exit{} \segment{s}, processes access shared variables pertaining to the lock (and the lock only). A process may crash at any point during its execution, including while executing \NCS, \Recover, \Enter, \CS{} or \Exit{} \segment{}.
        
        \begin{definition}[passage]
            A \emph{passage} of a process is defined as the sequence of steps executed by the process from when it begins executing \Recover{} \segment{} to either when it finishes executing the corresponding \Exit{} \segment{} or experiences a failure, whichever occurs first.
        \end{definition}
        
        
        \begin{definition}[super-passage]
            A \emph{super-passage} of a process is a maximal non-empty sequence of consecutive passages 
            executed by the process, where only the last passage of the process in the sequence is failure-free.
        \end{definition}
        

    \subsection{RME problem definition}
    \label{sec:problem}
    
        A \emph{history} is a collection of steps taken by processes. 
        A process $p$ is said to be \emph{active} in a history $H$ if $H$ contains at least one step by $p$.
        We assume that every critical section is finite.
        %
        A history $H$ is said to be \emph{fair} if 
        \begin{enumerate*}[label=(\alph*)]
            \item it is finite, or 
            \item if it is infinite and every active process in $H$ either executes infinitely many steps or stops taking steps after a failure-free passage.
        \end{enumerate*}
        %
        Designing a recoverable mutual exclusion (RME) algorithm involves designing \Recover, \Enter{} and \Exit{} \segment{s} such that the following correctness properties are satisfied.  
        
        \begin{description}
        
            \item[Mutual Exclusion (ME)] For any finite history $H$, at most one process is in its \CS{} at the end of $H$. 
        
            \item[Starvation Freedom (SF)] Let $H$ be an infinite fair history in which every process crashes only a finite number of times in each super passage. Then, if a process $p$ leaves the \NCS{} \segment{} in some step of $H$, then $p$ eventually enters its \CS{} \segment{}.
            
            \item[Bounded Critical Section Reentry (BCSR)] For any history $H$, if a process $p$ crashes inside its \CS{} \segment{}, then, until $p$ has reentered its \CS{} \segment{} at least once, any subsequent execution of \Enter{} \segment{} by $p$ either completes within a bounded number of $p$'s own steps or ends with $p$ crashing.
        \end{description}
        
        Note that mutual exclusion is a safety property, and starvation freedom is a liveness property. The bounded critical section reentry is a safety as well as a liveness property. If a process fails inside its \CS{}, then a shared object or resource (\emph{e.g.}, a shared data structure) may be left in an inconsistent state. The bounded critical section reentry property allows such a process to ``fix'' the shared resource before any other process can enter its \CS{} (\emph{e.g.}, \cite{GolRam:2019:DC,GolHen:2017:PODC,JayJay+:2019:PODC}). This property assumes that the \CS{} is idempotent; i.e, the \CS{} is designed so that, in a super passage, multiple executions of the \CS{} is equivalent to one execution of the \CS{}.

        Our correctness properties are the same as those used in \cite{GolRam:2019:DC,GolHen:2017:PODC,JayJay+:2019:PODC}. We have stated them here for the sake of completeness.
        In addition to the correctness properties, it is also desirable for an RME algorithm to satisfy the following additional properties.
        
        \begin{description}
        
            \item[Bounded Exit (BE)] For any infinite history $H$, any execution of the \Exit{} \segment{} by any process $p$ either completes in a bounded number of $p$'s own steps or ends with $p$ crashing.
         
            \item[Bounded Recovery (BR)] For any infinite history $H$, any execution of \Recover{} \segment{} by process $p$ either completes in a bounded number of $p$'s own steps or ends with $p$ crashing.
        
        \end{description}
        
    \subsection{Memory Reclamation problem definition}
        We only consider those RME algorithms that need to allocate nodes dynamically on the heap. We assume that the underlying RME algorithm needs to use a new node per request. A node is a collection of resources required by the underlying RME algorithm per request.
        
        The general memory reclamation problem involves designing two methods, 
        \begin{enumerate*}
            \item \textit{new\_node(~)}, and
            \item \textit{retire(node)}
        \end{enumerate*}. 
        These methods are used to allocate and deallocate nodes dynamically. The \textit{retire(node)} method assumes a node is retired only when there are no more references to it in shared memory, and no new shared references will be created. The responsibility of a memory reclamation service is to provide safe reclamation (defined later) of a node once it is retired. On the other hand, the responsibility of retiring the node is typically on the programmer that needs to consume the memory reclamation service.
        
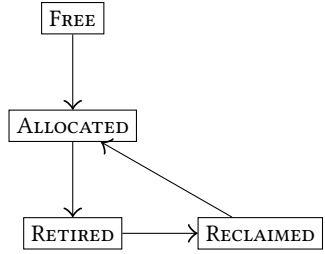
\begin{figure}[t]
        \scalebox{1.0}{
            \centering
        \begin{tikzpicture}[decoration={markings, mark=at position 1 with {\arrow[scale=2,black]{>}}}]]
            \node (F) [rectangle, draw]{\textsc{Free}};
            \node (A) [rectangle, draw, below=1cm of F]{\textsc{Allocated}};
            \node (R) [rectangle, draw, below=1cm of A] {\textsc{Retired}};
            \node (M) [rectangle, draw, right=1cm of R] {\textsc{Reclaimed}};
            
            \draw[postaction={decorate}] (F) -- (A);
            \draw[postaction={decorate}] (A) -- (R);
            \draw[postaction={decorate}] (R) -- (M);
            \draw[postaction={decorate}] (M) -- (A);
        \end{tikzpicture}
        }
        \captionof{figure}{The lifecycle of a node} 
        \label{fig:node_lifecycle}
    \end{figure}

        In our work, we assume that nodes are reused (instead of freed), once they are reclaimed. As a result, the lifecycle of a node follows four (logical) stages:
        \begin{enumerate*}
            \item \Freed{}
            \item \Allocated{}
            \item \Retired{}
            \item \Reclaimed{}
        \end{enumerate*}.
        The lifecycle of a node follows a pattern as shown in \autoref{fig:node_lifecycle}. Initially, a node is assumed to be in the \Freed{} stage. Once it is assigned by the \textit{new\_node(~)} method, it is in the \Allocated{} stage. After getting retired, it is in the \Retired{} stage, and finally, it is moved to the \Reclaimed{} stage by the memory reclamation algorithm. Once a node is reclaimed, it can be reused and will move to the \Allocated{} stage, and so on.
        
        Designing a memory reclamation scheme for recoverable mutual exclusion (RME) algorithms involves designing the \textit{new\_node(~)} and \textit{retire(node)} methods such that the following correctness properties are satisfied.
        
        \begin{description}
             \item[Safe reclamation] For any history $H$, if process $p_i$ accesses a node $x$, then either $x$ is local to $p_i$, or $x$ is in \Allocated{} or \Retired{} stages.
        \end{description}
        Note that any RME algorithm only requires a single node at any given point in time. Thus, we would want multiple executions of the new\_node(~) method to return the same node until the node is retired. Similarly, we want to allow the same node to be retired multiple times until a new node is requested. 
        \begin{description}
            \item [Idempotent allocation]
                Given any history $H$, process $p_i$ and a pair of operations, $op_1$ and $op_2$, of the \textit{new\_node(~)} method invoked by $p_i$, if there does not exist an invocation of \textit{retire(node)} by $p_i$ between $op_1$ and $op_2$, then either both these operations returned the same node in $H$, or at least one of these operations ended with a crash.
             
            \item [Idempotent retirement]
                Given any history $H$, process $p_i$ and a pair of operations, $op_1$ and $op_2$, of the \textit{retire(node)} method invoked by $p_i$, if there does not exist an invocation of \textit{new\_node(~)} by $p_i$ between $op_1$ and $op_2$, then either history $H' = H - \{op_1\}$ or $H'' = H - \{op_2\}$ or both are equivalent to $H$.
             
        \end{description}
        In case of failures, it is the responsibility of the underlying algorithm to detect if the failure occurred while executing any method of the memory reclamation code and if so, re-execute the same method.

    \subsection{Performance measures}
    
        We measure the performance of RME algorithms in terms of the number of \emph{remote memory references (RMRs)} incurred by the algorithm during a \emph{single} passage. Similarly, the performance of a memory reclamation algorithm for RME is measured in terms of RMR overhead per passage. The definition of a remote memory reference depends on the memory model implemented by the underlying hardware architecture. In particular, we consider the two most popular shared memory models:
        
       \begin{description}
            \item[Cache Coherent (CC)]
                The CC model assumes a centralized main memory. Each process has access to the central shared memory in addition to its local cache memory. The shared variables, when needed, are cached in the local memory. These variables may be invalidated if updated by another process. Reading from an invalidated variable causes a cache miss and requires the variable value to be fetched from the main memory. Similarly, write on shared variables is performed on the main memory. Under this model, a remote memory reference occurs each time there is a fetch operation from or a write operation to the main memory.
            \item[Distributed Shared Memory (DSM)]
                The DSM model has no centralized memory. Shared variables reside on individual process nodes. These variables may be accessed by processes either via the interconnect or a local memory read, depending on where the variable resides. Under this model, a remote memory reference occurs when a process needs to perform \textit{any} operation on a variable that does not reside in its own node's memory.
        \end{description}
    
        \subsection{Synchronization primitives}
        
            We assume that, in addition to read and write instructions, the system also supports \emph{compare-and-swap (\textit{CAS})} read-modify-write (RMW) instruction.
    
            A compare-and-swap instruction takes three arguments: $address$, $old$ and $new$; it compares the contents of a memory location ($address$) to a given value ($old$) and, only if they are the same, modifies the contents of that location to a given new value ($new$). It returns \textit{true} if the contents of the location were modified and \textit{false} otherwise.
    
            This instruction is commonly available in many modern processors such as Intel~64~\cite{Intel64Manual} and AMD64~\cite{AMD64Manual}.
    
    \section{The \broadcast{} object}
        \label{sec:broadcast}
        
        Our memory reclamation technique utilizes a recoverable \broadcast{} object whose primary function is to allow a designated process to signal (and free) multiple waiting processes. The \broadcast{} object is inspired by the SIGNAL object used by \JJJ{} \cite{JayJay+:2019:PODC} to solve the RME problem. Unlike the SIGNAL object, that can signal only one waiting process and perform signalling only once, the \broadcast{} object allows a designated process to signal multiple waiting processes and can be reused, even in the presence of failures.
        
        In essence, the \broadcast{} object is a recoverable MRSW (Multi Reader Single Writer) \textbf{counter} object that supports three operations \bSet{}, \bWait{} and \bRead{}.
        \begin{enumerate}[topsep=0ex]
            \item \bSet{($x$)} is invoked by a process to set the counter value to $x$
            \item \bWait{($x$)} is invoked by a process that intends to wait till the counter value is greater than or equal to $x$ 
            \item \bRead{(~)} is invoked to read the current value of the counter.
        \end{enumerate}  
        This object assumes the following usage:
		\begin{enumerate}[topsep=0ex]
			\item \bSet{} operation will only be invoked by a designated process $p_w$
			\item \bWait{} operation can be invoked by all processes except $p_w$
			\item \bSet{} operation must only be invoked in an incremental fashion. In other words, if the last successful \bSet{} operation was \bSet{$(y)$}, then the next invocation may only be \bSet{$(y+1)$} or, to maintain idempotence, \bSet{$(y)$}
			\item \bWait{} operation must not be invoked with a parameter whose value is at most one unit greater than the current counter value. Formally, if the last successful \bSet{} operation was \bSet{$(y)$}, and \bWait{$(z)$} is invoked, then $z \leq y+1$.
		\end{enumerate}
        
		An implementation of the \broadcast{} object is trivial in the CC model. Using a shared MRSW atomic integer, processes can achieve $\bigO{1}$ RMR-complexity for \bSet{}, \bWait{} and \bRead{}. However, this approach does not work for the DSM model. In the DSM model, each shared variable resides on a single processor node. Thus, if processes wait by spinning on the same shared variable, some processes (from remote nodes) will incur an unbounded number of RMRs. Thus, each process needs to spin on a variable stored in its local processor node. In this case, process $p_w$ needs to broadcast its \bSet{$(x)$} operation to ensure that all processes that are spinning due to an invocation of the \bWait{($x$)} operation are subsequently signalled. This action could potentially incur $\bigO{n}$ RMRs for the \bSet{$(x)$} operation. Thus, a constant-RMR implementation of the \broadcast{} object for the DSM model is non-trivial.
		
		We present an efficient implementation of the \broadcast{} object for the DSM model in \autoref{algo:Broadcast}. This implementation incurs $\bigO{1}$ RMRs for \bSet{}, \bWait{} and \bRead{} and utilizes $\bigO{n}$ space per \broadcast{} object. The main idea in our implementation of the \broadcast{} object is a wakeup chain, created by the designated process $p_w$, such that each process in the wakeup chain wakes up the next process in the chain. To trigger the wakeup in the wakeup chain, process $p_w$ only needs to wake up the first process in the wakeup chain.
        
		\begin{algorithm}[t]
			\begin{\algoFontSize}
				\begin{multicols}{2}
				
				\SetKw{Shared}{shared non-volatile variables}
				\SetKw{Local}{local variables}
				\SetKw{Struct}{struct}
				\SetKw{Integer}{int}
				\SetKw{Boolean}{bool}
				\SetKw{Array}{array}
				\SetKw{Await}{await}
				\SetKw{Writer}{Designated process}
				
				\Shared \\
				\Indp
				\tcc{Keep track of counter value}
				$count$ : atomic integer\;
				\tcc{Internal counter to synchronize \bSet{} and \bWait{}}
				$interim\_count$ : atomic integer\;
				\tcc{value to spin on; the $i$-{th} entry is local to process $p_i$} 
				$target[1{\dots}n]$: \Array $[1{\dots}n]$ of integer\;
				\tcc{announcement of target value to $p_w$; all entries are local to process $p_w$} 
				$announce[1{\dots}n]$: \Array $[1{\dots}n]$ of integer\;
				\tcc{id of next process in wakeup chain; all entries are local to process $p_w$}
				$wakeup[1{\dots}n]$: \Array $[1{\dots}n]$ of integer\;
				\Indm
				\BlankLine
				
				\vspace{1.2\baselineskip}
				
				\SetKwProg{fbWait}{Function}{}{end}
				\fbWait{\bWait{}(x)}
				{
					\tcc{Wait till counter value reaches $x$}
					\tcc{Process $p_w$ should never invoke \bWait{(x)}}
					$target[i] \leftarrow x$\;
					$announce[i] \leftarrow x$\;
					\tcc{No need to wait if $p_w$ intends to set counter to $x$}
					\If{$interim\_count \geq x$}
					{
						$target[i] \leftarrow 0$\;
					}
					\tcc{Spin till some process resets the target value}
					\textbf{\Await} $target[i] > 0$\;
					$announce[i] \leftarrow 0$\;
					$k \leftarrow wakeup[i]$\;
					\If{$k > 0$}
					{
						\tcc{Wake up next process in wakeup chain}
						CAS$(target[k], x, 0)$
					}
				}
				\BlankLine
				
				\SetKwProg{fbRead}{Function}{}{end}
				\fbRead{\bRead{}(~)}
				{
					\Return $count$
				}
				\BlankLine
				\columnbreak
				
				\Writer \\
				\Indp
				\tcc{Host process for \broadcast{} object}
				$p_w$: writer process\;
				\Indm
				\BlankLine
				
				\vspace{0.3\baselineskip}
				\SetKwBlock{DummyBlock}{}{}
				\SetKw{Initialization}{Initialization}
				
				\Initialization
				\SetAlgoNoLine\DummyBlock
				{
					\SetAlgoLined
					\tcc{counts are initially zero}
					$count \leftarrow 0$\;
					$interim\_count \leftarrow 0$\;
					\ForEach{$j \in \{ 1, 2, \dots, n\}$}
					{
						\tcc{processes are not waiting initially}
						$target[j] \leftarrow 0$\;
						$announce[j] \leftarrow 0$\;
						$wakeup[j] \leftarrow 0$\;
					}
				}\SetAlgoLined
				\BlankLine
				
				\vspace{0.5\baselineskip}
				\SetKwProg{fbSet}{Function}{}{end}
				\fbSet{\bSet{}(x)}
				{
					\tcc{Sets the counter value to $x$}
					\tcc{\bSet{(x)} may only be invoked by process $p_w$}
					$last \leftarrow 0$\;
					$j \leftarrow 1$\;
					\tcc{Inform all process about an incoming set operation}
					$interim\_count \leftarrow x$\;
					\While{j < n}
					{
						\If{$announce[j] = x$}
						{
						
							\tcc{Assign $j$ to wakeup the last waiting process}
							$wakeup[j] \leftarrow last$\;
							\If{$announce[j] = x$}
							{
								\tcc{$j$ is the last process if it is still waiting}
								$last \leftarrow j$\;
							}
						}
						$j \leftarrow j + 1$\;
					}
					\If{$last > 0$}
					{
						\tcc{Release the last process and all waiting processes will be automatically released using the wakeup chain}
						CAS$(target[last], x, 0)$\;
					}
					$count \leftarrow x$
				}
				\BlankLine
				
				\end{multicols}
				\caption{Pseudocode for \broadcast{} object for process $p_i$ for the DSM model}
				\label{algo:Broadcast}
			\end{\algoFontSize}
		\end{algorithm}
        
        \subsection{Variables used}
            The variable $count$ is used to store the counter value. The variable $interim\_count$ is used by \bSet{($x$)} to temporarily store the new counter value until the operation terminates. Variables $target$, $announce$ and $wakeup$ are arrays of integers with one entry for each process.
            The $i$-{th} entry of $target$ is used by process $p_i$ to spin on until the counter value of the \broadcast{} object reaches $target[i]$.
            The $i$-{th} entry of $announce$ is used by process $p_i$ to indicate process $p_w$ about its intention to wait for the counter value of the \broadcast{} object to be set to $announce[i]$.
            The $wakeup$ array is used to create a wakeup chain. The $i$-{th} entry of $wakeup$ is used by process $p_i$ to determine the next process in the wakeup chain. The $wakeup$ and $announce$ arrays are local to process $p_w$.
        
        \subsection{Algorithm description}
            In \bWait{($x$)}, a process $p_i$ first sets $target[i]$ and then $announce[i]$ to $x$, to announce its intention to wait for the counter value to reach $x$. It then checks if \bSet{} has been invoked for this particular value of $x$, by checking the variable $interim\_count$. If yes, it resets $target[i]$ to 0. It then spins, if required, till $target[i]$ is set to 0 by some process in the wakeup chain. Process $p_i$ then clears its announcement and determines the next process $p_k$ in the wakeup chain, where $wakeup[i] = k \neq 0$. It wakes up $p_k$ by updating the value of $target[k]$ from $x$ to $0$ using a CAS instruction. Note that there could be multiple wakeup chains in the algorithm for different $target$ values. However, the algorithm maintains the invariant that all processes in a particular wakeup chain have the same $target$ value.
            
            In \bSet{($x$)}, process $p_w$ first sets the $interim\_count$ to $x$ such that any process that invokes \bWait{} from this point does not get blocked. Then, it creates the wakeup chain of processes in a reverse order by keeping track of the last process in the wakeup chain and double checking the announce array to ensure the process is indeed waiting. Lastly, $p_w$ wakes up the first process in the wakeup chain (indicated by variable $last$) and subsequently all waiting processes will be woken up.
            
            The only use of the \bRead{(~)} operation is to keep track of the last successful \bSet operation. All \bSet{($x$)}, \bWait{($x$)} and \bRead{(~)} operations are idempotent methods and would execute perfectly even if run multiple times as long as they are run to completion once.
    
    \section{The memory reclamation algorithm}
        \label{sec:mem_rec}
        
        Our idea relies on the notion of a \textit{grace} period and quiescent states \cite{arcangeli2003using, mckenney1998read, hart2007performance}.
        \begin{definition}[Grace period]
            A grace period is a time interval $[a, b]$ such that all nodes retired before time $a$ are safe to be reclaimed after time $b$.
        \end{definition}
        \begin{definition}[Quiescent state]
            A process is said to be in a quiescent state at a certain point in time if it cannot access any node from another process using only its local variables.
        \end{definition}
        Note that quiescent states are defined within the context of an algorithm. Different algorithms may encompass different quiescent states. In the context of quiescent states, a \textit{grace} period is a time interval that overlaps with at least one quiescent state of each process. In order to reuse a node, a process, say $p_i$, must first retire its node and then wait for at least one complete \textit{grace} period to safely reclaim the node. After one complete grace period has elapsed, it is safe to assume that no process would be able to acquire any access to that node. 
        
        \emph{Main idea:} In the case of RME algorithms, we assume that when a process is in the \NCS{}, it is in a quiescent state. It suffices to say that after $p_i$ retires its node, if some process $p_j (j \neq i)$ is in the \NCS{} \segment{}, then $p_j$ would be unable to access that node thereafter. In order to safely reuse (reclaim) a node, process $p_i$ determines its grace period in two phases, the snapshot phase and the waiting phase. In the snapshot phase, $p_i$ takes a snapshot of the status of all processes and, in the waiting phase, $p_i$ waits till each process has been in the \NCS{} \segment{} at least once during or after its respective snapshot. In order to remove the RMR overhead caused by scanning through each process, $p_i$ executes each phase in a step manner.
        
        Our memory reclamation algorithm is provides two methods:
        \begin{enumerate*}[label=\arabic*)]
            \item \textit{\newnode{}(~)}, and
            \item \textit{\retire{}(~)}
        \end{enumerate*}.
        A pseudocode of the memory reclamation algorithm is presented in \autoref{algo:mem_rec}. Any RME algorithm that needs to dynamically allocate memory can utilize our memory reclamation algorithm by invoking these two methods.
        The \textit{\newnode{}(~)} method returns a ``node'' that is required by a process to enter the \CS{} of the RME algorithm. Similarly, while leaving the \CS{}, the \textit{\retire{}(~)} method will retire the node used to enter the \CS{}, 
        Our algorithm assumes (and relies on) the fact that each process \textbf{will} request a new node each time before entering the \CS{} and retire its node prior to entering the \NCS{} \segment{}. THE RMR overhead of our algorithm is $\bigO{1}$, while the space overhead is $\bigO{n^2 * sizeof(node)}$.
        
		\begin{algorithm}[t]
			\begin{\algoFontSize}
				\begin{multicols}{2}
				
				\DontPrintSemicolon
				
				\SetKw{Shared}{shared non-volatile variables}
				\SetKw{Local}{local non-volatile variables}
				\SetKw{Struct}{struct}
				\SetKw{Integer}{int}
				\SetKw{Boolean}{bool}
				\SetKw{Array}{array}
				\SetKw{Await}{await}
				
				\Shared \\
				\Indp
				\tcc{Counter of CS attempts; $i$-{th} entry is local to process $p_i$}
				$start[1 \dots n]$: \Array of integer\;
				\tcc{Broadcast object to wait for CS completion; $p_i$ is the writer for the $i$-{th} entry}
				$finish[1 \dots n]$: \Array of \broadcast{} object\;
				\Indm
				\BlankLine
				
				\Local\\
				\Indp
					\tcc{Array to store last observed value of start of other process}
					$snapshot[1 \dots n]:$ \Array of integers\;
					\tcc{Pool of nodes for memory management}
					$pool[0,1][1 \dots 2n+2]$: two pools of $2n+2$ nodes\;
					\tcc{Index of current pool}
					$currentpool$: integer\;
					\tcc{Index of backup pool}
					$backuppool$: integer\;
					\tcc{Counter to track steps taken since last pool switch}
					$index$: integer\;
				\Indm
				
				\BlankLine
				
				\SetKwBlock{DummyBlock}{}{}
				\SetKw{Initialization}{initialization}
				\Initialization
				\SetAlgoNoLine\DummyBlock
				{
					\SetAlgoLined
					\ForEach{$j \in \{ 1, 2, \dots, n\}$}
					{
						$start[j] \leftarrow 0$\;
					}
					\ForEach{$p \in \{p_1, p_2, \dots, p_n\}$}
					{
						$currentpool \leftarrow 0$\;
						$backuppool \leftarrow 1$\;
						$index \leftarrow 1$\;
					}
				}\SetAlgoLined
				\BlankLine
				
				\SetKwProg{func}{Function}{}{end}
				\func{\newnode{}(~)}
				{
					\label{line:mem_rec:newnode:begin}
					\If{$start[i] = finish[i]$}
					{
						\label{line:mem_rec:newnode:if}
						Step(~)\;
						\label{line:mem_rec:newnode:step}
						$start[i]++$\;
						\label{line:mem_rec:newnode:incstart}
					}
					\tcc{Return the node in $currentpool$ pointed by $index$}
					\Return $pool[currentpool][index]$\;
					\label{line:mem_rec:newnode:end}
				}
				\BlankLine
				\columnbreak
				
				\func{\retire{}(~)}
				{
					\label{line:mem_rec:retire:begin}
					\If{$start[i] \neq finish[i].\bRead{(~)}$}
					{
						\label{line_mem_rec:retire:if}
						$finish[i].\bSet{(start[i])}$\;
						\label{line:mem_rec:retire:incfinish}
					}
				}
				\BlankLine
				
				\func{Step(~)}
				{
					\tcc{$index$ will progress in each execution}
					\tcc{Blocking}
					\uIf{$index \leq n$}
					{
						\tcc{Take snapshot}
						$snapshot[index] = start[i]$\;
						\label{line:mem_rec:step:snapshot}
						$index++$\;
					}
					\uElseIf{$index > n$ \textbf{and} $index <= 2n$}
					{
						\tcc{Wait for others to finish doorway}
						\If{$index - n \neq i$}  
						{
							\tcc{No need to wait for self}
							$finish[index - n].\bWait{(snapshot[index - n])}$\;
							\label{line:mem_rec:step:waiting}
						}
						$index++$\;
					}
					\uElseIf{$index = 2n + 1$}
					{
						\tcc{Backup pool is now reliable}
						$currentpool \leftarrow backuppool$\;
						\label{line:mem_rec:step:pool_set}
						$index++$\;
					}
					\Else
					{
						\tcc{Reset backuppool}
						$backuppool \leftarrow 1 - currentpool$\;
						\label{line:mem_rec:step:pool_reset}
						\tcc{Reset index}
						$index \leftarrow 1$\;
						\label{line:mem_rec:step:index_reset}
					}
				}
				\BlankLine
				
				\end{multicols}
				\caption{Pseudocode for Memory reclamation for process $p_i$}
				\label{algo:mem_rec}
			\end{\algoFontSize}
		\end{algorithm}
        
        \subsection{Variables used}
            There are two types of non-volatile variables used in this algorithm, shared and local. Variables $start$ and $finish$ are shared non-volatile arrays of integers and \broadcast{} objects respectively, with one entry per process.
            The $i$-{th} entry of $start$ is used by process $p_i$ to indicate the number of new nodes requested.
            The $i$-{th} entry of $finish$ is used by process $p_i$ to indicate the number of nodes retired. The entries of $finish$ are \broadcast{} objects for other processes to spin on until $finish[i]$ exceeds a particular value.
            
            In addition, we use five local non-volatile variables. The variable $snapshot$ is an array of integers to take a snapshot of $start$ array. Variable $pool$ is a collection of $2 * (2n + 2)$ nodes that would be employed by the underlying mutual exclusion algorithm to enter into \CS{}. Variable $currentpool$ is used to keep track of the active nodes in $pool$ and $backuppool$ is used to account for failures while switching the value of $currentpool$. Variable $index$ is used to keep track of the number of times the method $Step(~)$ has been invoked since the last time $currentpool$ was switched.
            
        \subsection{Algorithm Description}
            Each process maintains two pools locally, reserve and active, each of $2n + 2$ nodes ($pool[0,1][1,\dots,2n+2]$). The reserve pool contains nodes that have previously been retired and are in the process of reclamation. The active pool contains a mix of reclaimed nodes that are ready for reuse, and retired nodes that were consumed from the active pool while trying to reclaim the nodes from the reserve pool. The retired and safe nodes in the active pool are separated by the local variable $index$.
            
            The $start$ and $finish$ counters function in sync and differ by at most one. If $start[i] - finish[i] = 1$ for some $i$, it implies that process $p_i$ has left the \NCS{}. On the other hand, if $start[i] - finish[i] = 0$, it implies that process $p_i$ is in the \NCS{} and in a quiescent state. In order to enter the \CS{}, a $p_i$ first requests for a new node by invoking the \textit{\newnode{}(~)} method (\autoref{line:mem_rec:newnode:begin}). This indicates that the process has left the \NCS{} and thus increments the $start[i]$ counter (\autoref{line:mem_rec:newnode:incstart}). Similarly, once a process needs to retire a node, it invokes the \textit{\retire{}} method (\autoref{line:mem_rec:retire:begin}) wherein it updates the $finish$ counter (\autoref{line:mem_rec:retire:incfinish}). The $start$ and $finish$ counters are guarded by if-blocks (\autoref{line:mem_rec:newnode:if}, \autoref{line_mem_rec:retire:if}) to warrant idempotence in case of multiple failures.
            
            A process can consume nodes from the active pool only after taking steps towards reclaiming nodes from the reserve pool (\autoref{line:mem_rec:newnode:step}). The memory reclamation steps are implemented in the $Step(~)$ method. The role of the $Step(~)$ method is two-fold. Firstly, it advances the local variable $index$ during each successful execution in order to guarantee a fresh node on every invocation of the \textit{\newnode{}(~)} method. Second, the $Step(~)$ method performs memory reclamation in three phases.
            \begin{enumerate}
                \item Snapshot (\autoref{line:mem_rec:step:snapshot}): $p_i$ takes a snapshot of $start[j]$ for all $j \in \{1,\dots,n\}$ 
                
                \item Waiting (\autoref{line:mem_rec:step:waiting}): $p_i$ waits for $finish[j]$ to ``catch up'' to $start[j]$ using a \broadcast{} object as described in \autoref{sec:broadcast}. Simply put, $p_i$ waits for \textit{very old} unsatisfied requests of other processes to be satisfied. In this context, a request is very old if $p_i$ overtook it $n$ times. This ensures that each process has been in a quiescent state before $p_i$ goes to the pool swapping phase
                
                \item Pool swap (\autoref{line:mem_rec:step:pool_set} and \autoref{line:mem_rec:step:pool_reset}): If process $p_i$ reaches this phase, it implies that at least one \textit{grace} period has elapsed since the nodes in the reserve pool were retired. At this point it is safe to reuse nodes from the reserve pool and $p_i$ simply swaps the active and reserve pool. In order to account for failures, this swap occurs over two invocations of the $Step(~)$ method and the $index$ variable is then reset (\autoref{line:mem_rec:step:index_reset}).
            \end{enumerate}
            
             Note that the algorithm is designed in such a way that multiple executions of the \textit{\newnode{}(~)} method will return the same node until the \textit{\retire{}} method is called and vice versa. This design aids to introduce idempotence in and accommodates the failure scenario where a process crashes before being able to capture the node returned by the \textit{\newnode{}(~)} method.

    \section{Applications}
        \label{sec:application}
        Golab and Ramaraju's algorithms \cite{GolRam:2016:PODC} have a bounded space complexity, but use the MCS algorithm as their base lock. The space complexity of the MCS algorithm may grow unboundedly. Using our memory reclamation algorithm, we can bound the space complexity of their algorithms.
        
        Two known sub-logarithmic RME algorithms, from Golab and Hendler \cite{GolHen:2017:PODC}, and, from \JJJ{} \cite{JayJay+:2019:PODC}, both use MCS queue-based structures. Memory reclamation in these algorithms is not trivial and requires careful analysis and proofs. Our memory reclamation algorithm fits perfectly with these algorithms. The main idea is to employ one instance of the memory reclamation algorithm at each level of the sub-logarithmic arbitration tree. As a result, the overall space complexity of these algorithms can be bounded by $\bigO{n^3}$.
        
        Dhoked and Mittal's algorithm \cite{DhoMit:2020:PODC}, also uses a MCS-queue based structure where the space complexity may grow unboundedly. Using a separate instance of our memory reclamation algorithm for each level of their adaptive algorithm, we can bound the space complexity of their algorithm to $\bigO{n^2 * \nicefrac{\log n}{\log\log n}}$
    
    \section{Related Work}
        \label{sec:related}
        \subsection{Memory reclamation}
        In \cite{Mic:2004:TPDS}, Michael used \textit{hazard pointers}, a wait-free technique for memory reclamation that only requires a bounded amount of space. Hazard pointers are special shared pointers that protect nodes from getting reclaimed. Such nodes can be safely accessed. Any node that is not protected by a hazard pointer is assumed to be safe to reclaim. Being shared pointers, hazard pointers are expensive to read and update.
        
        In \cite{Fra:2004:PhD}, Fraser devised a technique called epoch based reclamation (EBR). As the name suggests, the algorithm maintains an epoch counter $e$ and three limbo lists corresponding to epochs $e-1$, $e$ and $e+1$. The main idea is that nodes reitred in epoch $e-1$ are safe to be reclaimed in epoch $e+1$. This approach is not lock-free and a slow process may cause the size of the limbo lists to increase unboundedly.
        
        In \cite{mckenney1998read}, Mckenney and Slingwine present the RCU framework where they demonstrate the use of quiescent state based reclamation (QSBR). QSBR relies on detecting quiescent states and a grace period during which each thread passes through at least one quiescent state. Nodes retired before the grace period are safe to be reclaimed after the grace period. In \cite{arcangeli2003using}, Arcangeli et. al. make use of the RCU framework and QSBR reclamation for the System V IPC in the Linux kernel.
    
        In \cite{Bro:2015:PODC}, Brown presents DEBRA and DEBRA+ reclamation schemes. DEBRA is a distributed extension of EBR where each process maintains its individual limbo lists instead of shared limbo lists and epoch computation is performed incrementally. DEBRA+ relies on hardware assistance from the operating system to provide signalling in order to prohibit slow or stalled processes to access reclaimed memory.
        
        \subsection{Recoverable Mutual Exclusion}
        Golab and Ramaraju formally defined the RME problem in \cite{GolRam:2016:PODC}. They also presented four different RME algorithms---a 2-process RME algorithm and three $n$-process RME algorithms. The first algorithm is based on Yang and Anderson's lock \cite{YanAnd:1995:DC}, and is used as a building block to design an $n$-process RME algorithm. Both these RME algorithms use only read, write and comparison-based primitives. The worst-case RMR complexity of the 2-process algorithm is $\bigO{1}$ whereas that of the resultant $n$-process algorithm is $\bigO{\log n}$. Both RME algorithms have optimal RMR complexity because, as 
        shown in~\cite{AttHen+:2008:STOC, AndKim:2002:DC, YanAnd:1995:DC}, any mutual exclusion algorithm that uses only read, write and comparison-based primitives has worst-case RMR complexity of $\bigOmega{\log n}$. The remaining two algorithms are used as transformations which can be applied to the MCS algorithm. The third algorithm transforms the MCS algorithm to yield a constant RMR complexity in the absence of failures, but unbounded worst case RMR complexity. The fourth algorithm transforms the MCS algorithm to achieve bounded RMR complexity in the worst case.
        
        Later, Golab and Hendler \cite{GolHen:2017:PODC} proposed an RME algorithm with sub-logarithmic RMR complexity of $\bigO{\nicefrac{\log n}{\log \log n}}$ under the CC model using MCS queue based lock~\cite{MelSco:1991:TrCS} as a building block. This algorithm was later shown to be vulnerable to starvation~\cite{JayJay+:2019:PODC}. Ramaraju showed in~\cite{Ram:2015:Thesis} that it is possible to design an RME algorithm with $\bigO{1}$ RMR complexity provided the hardware provides a special RMW instruction to swap the contents of two arbitrary locations in memory atomically. Unfortunately,  at present, no hardware supports such an instruction to our knowledge.
        
        In~\cite{JayJos:2017:DISC}, Jayanti and Joshi presented a fair RME algorithm with $\bigO{\log{n}}$ RMR complexity. Their algorithm satisfies bounded (wait-free) exit and FCFS (first-come-first-served) properties and only requires a bounded amount of space consumption. In~\cite{JayJay+:2019:PODC}, \JJJ{}  proposed an RME algorithm 
        that uses MCS queue-based structure to achieve sub-logarithmic RMR complexity of $\bigO{\nicefrac{\log n}{\log \log n}}$. To our knowledge, this is the best known RME algorithm  as far as the worst-case RMR complexity is concerned that also satisfies bounded recovery and bounded exit properties.
        
        In \cite{DhoMit:2020:PODC}, Dhoked and Mittal use the MCS queue-based lock to present an adaptive transformation to any RME algorithm whose RMR complexity is constant in the absence of failures and gradually adapts to the number of failures. The RMR complexity of their algorithm is given by $min\{\sqrt{F}, \nicefrac{\log n}{\log\log n}\}$. Using a weaker version of starvation freedom, Chan and Woelfel \cite{ChaWoe:2020:PODC} present a novel solution to the RME problem that incurs a constant number of RMRs in the amortized case, but its worst case RMR complexity may be unbounded.
       
        In~\cite{GolHen:2018:PODC}, Golab and Hendler proposed an RME algorithm under the assumption of system-wide failure (all processes fail and restart) with $\bigO{1}$ RMR complexity.
    
    \section{Conclusion and Future work}
        \label{sec:concl_future}
        In this work, we formalized the problem of memory reclamation for recoverable mutual exclusion algorithms and present a plug-and-play solution that can be used by existing and new RME algorithms. Our algorithm is RMR-optimal for both the CC and DSM models. Next steps would be to design a recoverable memory reclamation for RME that satisfies some notion of fairness. Another direction of work involves formulating the problem of memory reclamation for recoverable lock-free data structures and designing algorithms for the same.
    \bibliographystyle{ACM-Reference-Format}
    \bibliography{Citations}
\end{document}